\documentclass[10pt]{article}
\usepackage{thumbpdf}
\usepackage{color}
\usepackage[utf8]{inputenc}
\usepackage{authblk}
\usepackage{color}
\usepackage{graphicx, subfigure}
\usepackage{amsmath}

\usepackage{amsmath}
\usepackage{amssymb}
\usepackage{graphicx}
\usepackage{setspace}
\usepackage[utf8]{inputenc} 
\usepackage[T1]{fontenc}
\usepackage{eepic}
\usepackage{cite}
\usepackage{url}
\usepackage{epstopdf}
\usepackage{hyperref}
\usepackage[labelfont=bf,labelsep=period,justification=raggedright]{caption}

\makeatletter
\renewcommand{\@biblabel}[1]{\quad#1.}

\renewcommand{\subsubsection}[1]{\vskip 1em \noindent\textit{#1}.}

\newcommand{\beq}{\begin{equation}}
\newcommand{\eeq}{\end{equation}}
\newcommand{\beqa}{\begin{eqnarray}}
\newcommand{\eeqa}{\end{eqnarray}}

\makeatother

\date{}

\begin{document}

\begin{flushleft}
{\Large
\textbf{Reduction of systemic risk by means of Pigouvian taxation}}

Vinko Zlati\'{c}$^{1,\ast}$
Giampaolo Gabbi$^{2,3}$
Hrvoje Abraham$^{4}$
\\
\bf{1} Theoretical Physics Division, Rudjer Bo\v{s}kovi\'c Institute, P.O.Box 180, HR-10002 Zagreb, Croatia
\\
\bf{2} University of Siena, Department of Management and Law, Piazza S. Francesco, 53100 Siena, Italy
\\
\bf{3} SDA Bocconi School of Management, via Bocconi 8, 20136 Milan, Italy
\\
\bf{4}Artes Calculi, Deren\v{c}inova 1, HR-10002 Zagreb, Croatia
\\
$\ast$ E-mail: vinko.zlatic@irb.hr
\end{flushleft}

\section*{Abstract}

We analyze the possibility of reduction of systemic risk in financial markets through Pigouvian taxation of financial institutions which is used to support the rescue fund. We introduce the concept of the cascade risk with a clear operational definition as a subclass and a network related measure of the systemic risk. Using financial networks
constructed from real Italian money market data and using realistic parameters, we show that the cascade risk can be substantially reduced by a small rate of taxation and by means of a simple strategy of the money transfer from the rescue fund to interbanking market subjects. Furthermore, we show that while negative effects on the return on investment ($ROI$) are direct and certain, an overall positive effect on risk adjusted return on investments ($ROI^{RA}$) is visible. Please note that \emph{the taxation} is introduced as a monetary/regulatory, not as a fiscal measure, as the term could suggest. \emph{The rescue fund} is implemented in a form of a common reserve fund.

\section*{Introduction}

Since the onset of the global financial and economic crisis there has been a lot of suggestions on how to prevent similar future turmoils, how to minimize the contagion likelihood and how to manage the trade-off between stability and ineficiency of the financial system~\cite{angelini1996systemic, eisenberg2001systemic, iori2006systemic, may2008complex, haldane2011systemic, cont2011network}. One of the most severe issues that regulators and think-tanks encountered to assess the systemic risk was a lack of reliable data~\cite{Whelan2009}. Microeconomic data with lending patterns are particularly scarce and the analyses of the money market (MM) stability with respect to interconnectedness of the system suffers with the issue to be adequately tested with real observations. The systemic risk as a consequence of interconnectivity of market subjects was marked as a clear cause for the wide spread of the crisis~\cite{haldane2011systemic, cont2011network, Upper2011111}. At the peak of the crisis, while central banks 
and regulatory bodies struggled to 
come up with short and long term refinancing facilities and/or quantitative easing solutions, it was clear that there are no metrics or models on how to manage negative implications of interconnectivity of market players. Not only was it not clear how to stabilize the system, but there was not a single available metric how to clearly manage any aspect of interconnectivity and its consequences.

There has been a substantial rise in systemic risk research, especially since 2008~\cite{Upper2011111}. Part of an effort targeted qualitative analysis of potential contagion spreading while others focused on quantifying the systemic risk of nodes. Relation between systemic risk and networks has drawn more attention~\cite{angelini1996systemic, eisenberg2001systemic, iori2006systemic, may2008complex, haldane2011systemic, cont2011network, freixas2000systemic, iori2001criticality, boss2004network}.

One of the efforts to face the lack of reliable metrics to estimate the systemic risk exposure was the proposal of the DebtRank algorithm~\cite{battiston2012debtrank}, a systemic risk metric designed to measure the impact of default of some subject on the entire system. This measure was thoroughly investigated by a few central banks and we decided to use it as a proxy process for evaluation of cascade risk.

Aside DebtRank, during the last few years, other systemic risk measures were constructed as well. The most popular are Marginal Expected Shortfall (MES)~\cite{acharya2012measuring}, Systemic Risk Measure (SRISK)~\cite{brownlees2010volatility}, Delta Conditional Value-at-Risk (DCoVaR)~\cite{adrian2011covar}, Link-Aware Systemic Estimation of Risks (LASER)~\cite{Hu2012}, and PA \& GCA measures~\cite{drehmann2013measuring}. Only some of these metrics (DCoVaR, LASER and PA \& GCA) are network-based models. Moreover, most of them do not provide an analytical methodology.

Furthermore, there is a number of papers devoted to the network structure as promoter/inhibitor of different contagion processes related to the default cascade~\cite{d2012robustness, battiston2012liaisons, battiston2012default}. These more theoretical approaches point to the very strong dependence of the generic contagion spreading on the details of the network structure.

The classification of contagion suggested by the World Bank distinguishes between three definitions (broad, restrictive and very restrictive) of contagion on the basis of the nature of its occurrence. According to the broad definition, contagion is referred to as a “cross-country transmission of shocks or general cross-country spillover effects”. This definition assumes that contagion can be caused by any type of linkages between countries (fundamental and non-fundamental). According to the second, more restrictive, definition “contagion is a transmission of shocks to other countries beyond any fundamental links or common shocks”. Contagion in this sense can be caused by “irrational” phenomena (e.g., financial panic, herd behavior). According to the third, very restrictive, definition, financial contagion does not involve fundamental linkages either. It assumes that “contagion occurs when cross-country correlations increase during “crisis times” relative to correlations during “tranquil times”. This 
definition allows a feasible econometric measurement of contagion. Within the money market, systemic risk can be associated with different feature of its microstructure: default cascades, freezing of the money flow, herding behavior and many possible unknown unknowns.

In this paper we focus on the the overall systemic risk that we name \emph{cascade risk}. We argue that the cascade risk is a good measure of the systemic risk and clearly define conditions for applicability of our approach. For this purpose we use the following operational definition of cascade risk: 

\begin{quotation}
The cascade risk of the system is the probability that a randomly chosen financial institution will default due to a contagion propagating through a financial system initiated by the default of some other institution.
\end{quotation}

Clearly this definition does not take into account the impact on the real economy and the capital loss in the whole system, and it does not consider the increase of the individual probability of default. Nevertheless, it clearifies the process we focus on within the set of all the possible processes that constitute systemic risk.

We will find out and quantify the potential damage of systemic shocks to a well defined network of financial institutions. Also, we will discuss critical values of parameters for which one can expect to cover most of the risk. At the same time, policy implications of our proposal are consistent with the existing regulatory framework.

Past research proposed several measures for the reduction of systemic risk. One class of them include rising the capital requirements of market subjects~\cite{Horvath2013, gauthier2012macroprudential, hart2011new, hellwig2010capital}. Although increasing the quantity and quality of bank capital allows to reduce the exposure to some relevant risks, such as credit and market, it could be highly expensive when introduced in period of crisis. Another weakness of the capital increase is that banks are expected to pay the cost of capital creating a vicious cycle which concludes into a a higher financialization level, an unsustainable speculative exposure and, finally, a more probable systemic risk. The capital requirement for systemic financial institutions, as proposed by the Basel Committe after the crisis (Basel III framework) could also alter the purpose to guarantee a levelled playing field among competitors. In this approach every subject should rise enough capital to stop the contagion in its neighborhood 
so that the capital is available if contagion starts in the very proximity of any of the subjects.

Others propose a money market (MM) transactions tax~\cite{Poledna2014, darvas2011financial, schafer2012financial}. One can tax all transactions or only transactions which increase systemic risk. We argue that this is not a viable approach because of two main reasons. First, taxing all transactions needlessly increases the cost of an entire MM segment even if it is in the process of reducing systemic risk. This approach would discourage MM activity although its main purpose is to reduce liquidity fluctuations in the network in a very short period of time, which is often measured in minutes. Applying a tax on MM transactions would affect reciprocity when trades compensate monthly, weekly or even daily. Taxing one transaction in a zero netted pair would just introduce pointless additional costs. Taxing only transactions which contribute to increase of systemic risk is in a similar category since it does not account for the transaction net effect over a period of time. For instance, a pair of zero netted 
transactions would be taxed despite the fact they have no effect on market stability. This approach would also discourage unconstrained MM activity with effects on market liquidity which are not discussed. Additionally, identifying transactions which increase systemic risk is an operational challenge. The implementation of the tax proposal appears to be tricky for the impossibility to identify in real time trades and allow financial agents to estimate the net fiscal impact on profitability. The effect would be to freeze the market and, consequently, the liquidity sources could rapidly evaporate. MM venues would seize to exist in the present form, market-making would be virtually impossible to implement, and it could all badly end up with the introduction of the quotation matrix.

The questions this paper tries to answer can be addressed as follows: How much a small portion of balance sheet reserves allocated to the rescue fund increases overall financial stability of the system? Is there any possible gain for the institutions subject to this regulation that could outweigh the decrease of market profitability as a consequence of the proposed taxation?

We simulate the effect of Pigouvian taxation and the creation of the rescue fund aimed at reducing the cascade risk and, consequently, the overall systemic risk. Due to the lack of the real world data, we had to assume certain simplifications and idealizations to prove the concept. We do not provide quantitative solutions for the financial system stability. Instead, our purpose is to provide a framework which clearly demonstrates that the Pigouvian taxation can in principle greatly reduce the cascade and the overall systemic risks. We focus on the framework which can be readily used to quantify the cascade risk and effects of Pigouvian taxation and the usage of the rescue fund. We also demonstrate that, in a specific realistic scenarios, the rescue fund can increase the risk adjusted return on investment ($ROI^{RA}$). We show that the rescue fund in our setting:
\begin{enumerate}
\item reduces systemic risk for realistic levels of reserve requirement;
\item increases the risk adjusted ROI for individual financial institutions and for the financial market in general;
\item in certain scenarios can assure the same level of stability for significantly reduced rates of reserve requirements;
\item produce qualitatively the same results in all the cases of the networks we report here.
\end{enumerate}

In this paper we apply the DebtRank~\cite{battiston2012debtrank} to model the cascade risk, and we use real European e-MID overnight money market data to calibrate the networks among agents~\cite{iori2008network, finger2012network, raddant2013structure, hatzopoulos2013quantifying}. Pigouvian taxation is then used to finance \emph{the rescue fund}, which is used in the case of default of single financial players.

The paper is organized as follows. First, we describe the data and the additional parameters we use to originate financial networks. Second, we give an overview of the DebtRank methodology. Third, we propose the taxing mechanism and the strategy for the application of the rescue fund. Fourth, we explain the way we compute the systemic risk and $ROI^{RA}$. Fifth, we discuss our results and implications of the proposed taxation on market profitability and market stability in general. 

\section*{Data}

The model introduced in this research is based on a simplified financial agents' network, interconnected via mutual lending and borrowing activities. To calibrate financial assets and liabilities we use the real European e-MID overnight money market data. This is the only electronic market for interbank deposits in the Euro area and the USA. The data base is composed by the records of all transactions registered in the period from 1999 to 2009. Based on the body of work done in the field of complex networks we know that:

\begin{enumerate}
\item Real financial networks are very heterogeneous both in node degree and in node strength in all previously reported cases~\cite{iori2008network, finger2012network, raddant2013structure, hatzopoulos2013quantifying}.
\item Real financial networks exhibit a strong core-periphery structure~\cite{hojman2008core, fricke2012core}.
\item Small ($< 10,000$ nodes) and dense weighted networks, such as financial networks, are constrained in configuration space and should be similar to each other~\cite{zlatic2009rich}.
\end{enumerate}

Above mentioned facts support the use of e-MID data, since all previously reported facts suggest that e-MID is sufficiently representative for a broad class of financial networks. Clearly the difference between the sizes of, for example, U.S. and European financial systems are large, and although we could expect that the presented results would be quantitatively very different, we expect that qualitatively they should be very similar.

For the purpose of the paper we present $4$ different networks taken in $4$ different periods with respect to the overall economic conditions. The networks are aggregate of one month of transactions for: April 2000, October 2004, November 2008 and September 2009. We choose these periods to compare $2$ periods before and $2$ periods after the great financial crisis. 

Unfortunately, the field of complex networks is still missing a good reliable model for creation of weighted directed networks with strong heterogeneity. Presented research would greatly benefit if such model would be available. We could generate different realizations of financial networks to create a large number of idealized realizations. This way we trade off a more detailed understanding of correlations between results and imposed network parameters for the model of taxation on networks which are as realistic as possible. 

The networks we use are relatively small, ranging in size from $N=112 $ to $N=177$ nodes. Their out and in degree are presented in figure~\ref{Fig: NetworkProperties} panels a) and b). Although small, these networks exhibit substantial heterogeneity in their connection patterns. This is a ubiquitous feature of all investigated financial networks. In the same figure on the panels c) and d) we present out and in network strength. Heterogeneity is again obvious in both cases. An example of one network used in simulations is presented in figure~\ref{Fig: Network}.

\section*{Methods}

\subsection*{Parameters}
In order to compute the cascade risk using the DebtRank methodology we impose some assumptions with respect to bank parameters. First, we define reserve requirements and then we simulate different levels of their ratios.

Assets and liabilities of subject $i$ consist of five characteristic parts: \emph{ total balance} $B_{i}$, \emph{ assets lent} to interbanking market network subject $j$ $A_{ij}$, \emph{ liabilities borrowed} $A_{ji}$ from network subject $j$, \emph{ regulatory required balance sheet reserve} $E_{i}$ and \emph{ regulatory required rescue fund} $F_{i}$, held in the rescue fund. Parameters found in e-MID data are interbanking market lending $A_{ij}$, and interbanking market borrowing $A_{ji}$. Total assets lent to other participants of the trading network is $S^L_i=\sum_{j}A_{ij}$ and the total assets borrowed from other participants is $S^B_i=\sum_{j}A_{ji}$. We use this two variables as out and in strength of the weighted network. We simulate balances as follows:
\beq
\label{eq: Balance}
B_{i}=\beta*max(S^B_i,S^L_i).
\eeq
Parameter $\beta$ is estimated through the publicly presented data of the money market activity compared to the total banking assets volume for the European region~\cite{Schluter2012}. The typical value used in the paper is $\beta=10$ although we experimented with values ranging from $\beta\in[4,20]$. To calibrate assets and liabilities using MM data exposures, we a simple strategy of assigning the same $\beta$ to all market participants. Factor $\beta$ can be interpreted as an average of variable $\beta_i=B_{i}/max(S^B_i,S^L_i)$ computed separately for each institution. As long as the average $\beta$ is a good representative of typical value of the variable $\beta_i$, our estimates are economically and statistically robust.

Reserve requirements $E_i$ are also kept in constant ratio to balance.
\beq
\label{eq: Reserve}
E_{i}=\eta*B_i.
\eeq
We report many different values of $\eta\in[0,0.05]$ as reserve requirements vary largely depending on country and economic region~\cite{o2007reserve}. Reserve requirements are regulatory attempt to force market participants to always retain some degree of resilience to market shocks and to manage the deposit multiplier for banks. In the simplified systemic liquidity crunch we assume that if the bank short term obligations become greater then its reserves - the bank would fail. Realistically reserves can be very different from legally required minimum. In this case we can again assume that each institution has a reserve fraction $\eta_i=E_i/B_i$ and that the distribution of these fractions is well represented by their average value $\eta$.

The last parameter designed in our paper is the $i$ reserve fund obligation $F_i$ (for any agent i) affecting the market stability. We propose that each bank puts an amount
\beq
\label{eq: Fund}
F_i=\alpha*E_i
\eeq
in the fund, $\alpha\in[0,1]$. Therefore we use a simple strategy in which banks participate in the reserve fund proportionally to their size. Clearly different strategies in the real world are possible - a flat amount tax, for instance. Other interesting choice would be to use some metric, like DebtRank, to asses banks importance to the stability of the system and to tax proportionally to that metric. We decided to use proportional tax as we find it to be the most logical to minimize regulatory inequalities and arbitrage opportunities.

\subsection*{DebtRank}

A previously mentioned level of default $E_i$ is assigned to each node $i$ of the weighted directed network. This level represents a threshold for the bank default, i.e. if the sum of noncollectable loans passes this threshold the subject fails. The \emph{ initial level of distress} $\psi_i\in[0,1]$ is also assigned to each node $i$. \emph{The level of distress} $h_i(t)$ with $h_i(t)\geq \psi_i$ is computed as an amount of losses suffered by the bank $i$ after an iteration $t$ of the model, divided by its own level of default $E_i$. The DebtRank starts from the initial set of nodes whose $\psi_i>0$. Loans are represented with outgoing weighted links and each node $i$ at iteration $t$ increases a level of distress $h_j(t)=min\left\{1,h_j(t-1)+W_{ij}h_i(t-1)\right\}$ to its neighbor $j$. Here $W_{ij}=\frac{A_{ij}}{E_j}$ is a fraction of distress that can pass through the given link. Every link in the simulation can be used only once and we assign a label $l\in\{0,1\}$ to it. At time 0 all the links have label 
$l=0$ and after they are used they get label $l=1$ and are discarded from future consideration. The algorithm works in steps and all links have to be used before the next iteration. The last run is experienced when there are no more available links that could propagate the stress further. Finishing levels of distress $h_i$ give the impact of the process on financial system.

In our simulations we start with a default of one agent $s$, i.e. $\psi_s=1$. We do not add additional stress to other market subjects as this has one to one correspondence with lowering $E_i$ for the same amount of stress. In realistic setting with additional data it would make sense to simulate additional levels of shock, if one would be interested in difference between the expected and the real systemic risk. Some other authors simulate and describe the dynamics of risk propagation process \cite{RIS_0} and thus have to take into account subjects' correlations via mutual asset classes. We instead use the DebtRank procedure to estimate the influence of every node to the others and estimate their systemic importance. In this procedure subjects are connected only via their mutual financial dependencies. Timescale of the process is assumed small compared to the timescale in which correlations among asset clases become important. Additionally at our timescales, correlations can be represented as different 
initial probabilities of defaults for each node.

\subsection*{Rescue fund strategy}

When the first bak defaults, we simulate cascading effects on other interbanking market counterparts. The strategy of the rescue fund operation we simulate is the following. First the initial subject $i$ upon request gets its fund amount $F_{i}$ to try to prevent its default. We assume that the initial subject always defaults. The rationale for such a decision is as follows: since every subject can ask for its own share from rescue fund if it is needed, the reserve level $E_{i}$ is constant before the default cascade. This means that the probability the first bank fails does not depend on the size of the rescue fund.

One of the issues for financial regulators is to minimize the moral hazard risk when bankers believe to be rescued with bail-outs solutions. In our model we introduce the opportunity to manage a default without a serious contagion, reducing the systemic risk.

When the first subject defaults all of its neighbors (lenders) receive all the money needed to stop further catastrophic cascading effects up to a value of the complete rescue fund. This is the Pigouvian idea devised in a way that no participant is subject to moral hazard of risky behavior under the protection of rescue fund, but all other market participants know that they are protected with respect to risky behavior of their counter parties. If the fund is not big enough to cover all the loses the money is distributed proportionally to the sum each subject has requested.

To make this strategy viable it is \emph{a priori} desirable that:
\begin{enumerate}
 \item the size of the rescue fund is big enough to cover the loss of financial intermediaries in the market, particularly network hubs, as a consequence of some subject default;
 \item the size of the rescue fund is small enough to be easily and urgently replaced with the fresh liquidity once the fund has been used;
 \item the implemented procedures of the rescue fund enable momentary usage of the fund.
\end{enumerate}

\subsection*{Cascade risk}

Here we finally present the cascade risk and the method we use to estimate it. Since we defined it a an measure of the probability of default, we address the probability of default of subject $i$ in a given time horizon. We split the probability of default into two components. The first component is exogenous and it is caused by an external market event $p_{i}^{O}$; the second component, independent from the first factor, is endogenous and it caused by interbanking systemic network related risk $p_{i}^{S}$:
\beq p_{i} = p_{i}^{O}+p_i^{S}.\eeq
Here we operate with a simple assumption that $p_i^{O}$ of every bank is small enough that we can expect only one default in the monitored period for which we estimate systemic risk, i.e. $max(p_i^{O})*N\ll 1$. We omit the ''set intersection'' term $- p_{i}^{O}p_i^{S}$ assuming that it is small compared to the linear approximation. This assumption does not change qualitative behavior of our taxation analysis but introduces only a slight quantitative changes. Here $N$ is the number of subjects in the network. For the endogenous probability of default $p_{i}^{S}$ we can write:
\beq\label{eq: exogenous}
p_{i}^{S}=\sum_{j\neq i} Q(i|j)p_{j}^{O},
\eeq
where $ Q(i|j)$ is conditional probability that subject $i$ will default given that subject $j$ has defaulted due to exogenous factors. Notice that if we would accept the possibility of simultaneous events, the real exogenous probability would be:
\beq\label{eq: exogenousReal}
p_{i}^{S}=\sum_{s\geq 1}^{N-1}\sum_{\substack{\Omega(s)\subset\\ \Omega(N-1)}} Q(i|\Omega(s))\prod_{j\in\Omega(s)}p_{j}\prod_{j'\notin\Omega(s)}\left(1-p_{j'}\right)
\eeq
where $s$ is the number of subjects that simultaneously default, $\Omega(s)$ is the set of subjects which default simultaneously and his size is $s$, and $ Q(i|\Omega(s))$ is the conditional probability that subject $i$ will default if all subjects in the set $\Omega(s)$ default simultaneously. The first sum in~(\ref{eq: exogenousReal}) runs over all possible $s$ - a number of subjects which could default simultaneously thus igniting a default cascade. The second sum runs over all the possible sets which contain the same number of subjects $s$. Products weight the probability of occurrence of every given set of initially defaulting subjects under assumption that these events are \emph{not} correlated to each others.

We use the definition of the cascade risk $p_{i}^{C}$, mentioned before and write within the approximation of only one defaulting subject:
\beq
\label{eq: CascadeRisk}
p_{i}^{C}=\frac{\sum_{j\neq i} Q(i|j)p_{j}^{O}}{\sum_{j\neq i}p_{j}^{O}}.
\eeq

The cascade risk of subject $i$ is the conditional probability that the cascade will reach the subject $i$ not depending on subject $j$ which initiated the cascade. Therefore we weight all the paths by probability calibrating proportionally to $p_{j}^{O}$, and we can label every cascade path by its starting node $j$. An important property of this measure is that it is invariant to scaling of exogenous probabilities, i.e. if we scale all the exogenous probabilities for the same amount $\gamma p_j^{O}\rightarrow p_j'^{O} \Rightarrow p_{i}'^{C}=p_{i}^{C}$. This means that the values of exogenous probabilities are neutral. 

We adopt the DebtRank algorithm for the simulation of default cascade, the conditional probbaility that the $j$ default will affect the $i$ default $ Q(i|j)$ will be coherently computed. Notice that DebtRank always provides the same solution for the same set of initial conditions, so the $ Q(i|j)$ is modelled as a binomial variable. If we further simplify our assumptions and suppose that exogenous probabilities of the default are all the same, i.e. that there is no difference among the subjects, the equation~(\ref{eq: exogenous}) reduces to
\beq\label{eq: exogenous DebtRank}
p_{i}^{S}=\Delta_ip^{O},
\eeq
where $\Delta_i$ is the number of times the subject $i$ has defaulted during the algorithm execution, and $p^O$ is exogenous probability of default which is the same for all the subjects. In this case overall probability of default is
\beq\label{eq: Final p}
p_{i} = \left(1+\Delta_i\right)p^{O}.
\eeq
The cascade risk of subject $i$ in this case is:
\beq
\label{eq: CascadeRiskDR}
p_{i}^{C}=\frac{\Delta_i}{N-1}.
\eeq
The cascade risk of the system we define as the average cascade risk of all subjects:
\beq
\label{eq: AverageCascadeRiskDR}
p^{C}=\frac{\sum_i p_{i}^{C}}{N}=\frac{\sum_i\Delta_i}{N(N-1)}.
\eeq

(11) is the measure that quantifies individual exposure of the institution to risky behavior of other financial counterparts. Notice that the cascade risk does not explain anything about the number of subjects that can default or about the level of financial shock the MM will suffer in the case of default cascade. It should therefore be seen as an individual measure to explain the risk that the financial system estimates for the individual subject. 

\subsection*{Effect on the return on investment (ROI)}

In order to devise a scheme for reduction of systemic risk with realistic chances for implementation it should minimize the conflicts among all the different stakeholders involved in the equilibrium process. We believe that a scheme that can greatly reduce cascade risk, and therefore systemic risk could also have a positive effect on the overall return on investment of financial intermediaries. To reduce and maximally simplify the model, we investigate the ROI of banks. If the effective risk adjusted return on investment of the market participant's assets is significantly increased through the positive effects of the rescue fund taxation, there could also be positive effects to their debt cost along with the market acceptance of their securities, with a virtuous circle. To quantify this effect we use the following method.

Every subject records a \emph{nominal} return of investment of their assets $ROI_{i}^{N}$ and a \emph{risk-adjusted} counterpart $ROI_{i}^{RA}$.

We classify the ROI in two categories: the $ROI^N$ for assets inside the network is $ROI^{int}$, and the ROI for assets outside the network is $ROI^{ext}$. We also introduce $ROI^E$ which is the return on investment tied to the balance sheet reserve. In that case the total nominal return on investment is

\beq\label{eq: ROI^N}
ROI_{i}^N = \frac{ROI^{int}\,\sum_{j\neq i} A_{ij} + ROI^{ext}D_i+ ROI^E E_i}{B_i},
\eeq

where $D_i=B_i-\sum_{j\neq i} A_{ij}-E_i$ is a total volume of assets allocated outside the money market.

\emph{Risk-adjusted} ROI is defined as
\beq\label{eq: ROI^RA}
ROI_{i}^{RA} = ROI_{i}^{N}\, (1-p_{i}) - p_{i},
\eeq
where $p_{i}$ is the bank $i$ investment horizon default probability.
This enables us to calculate the effects of rescue fund on risk-adjusted return on investment $ROI^{RA}$. Before the fund introduction we can compute $ROI^{RA}$ as follows:
\beqa\label{eq: ROI^RA2}
ROI_{i}^{RA} &=& ROI_{i}^{N}\, \left(1- p_{i}^{O}-p_i^{S}\right) -  p_{i}^{O}-p_i^{S}\nonumber\\
&=& ROI_{i}^{N}\,\left(1-p_{i}^{O}-\sum_{j\neq i} Q(i|j)p_{j}^{O}\right) -  p_{i}^{O}-\sum_{j\neq i}Q(i|j)p_{j}^{O}.
\eeqa
Using the assumption that exogenous probabilities are all the same, i.e. $\forall j:~p_j^{O}=p^{O}$, we obtain:
\beq\label{eq: ROI^RA DR}
ROI_{i}^{RA} = ROI_{i}^{N}\, \left(1- (1+\Delta_i) p^{O}\right) - \left(1+\Delta_i\right)p^O.
\eeq

After introducing the fund parametrized through the taxation rate $\alpha$ of the balance sheet reserves, default probabilities change to $p_{i}(\alpha)$ and systemic probabilities become $p_{i}^{S}(\alpha)$. Note that exogenous default probabilities $p_{i}^{O}$ and $p^O$ do not change as our strategy allows for every subject to withdraw the money from the fund if needed. Now the risk-adjusted ROI becomes

\beq\label{eq: ROI^RA alpha}
ROI_{i}^{RA}(\alpha) = ROI_{i}^{N}(\alpha)\, \left(1-p_{i}(\alpha)\right) - p_{i}(\alpha).
\eeq
Nominal $ROI_{i}^{N}(\alpha)$ also depends on $\alpha$ because some portion of assets is allocated to the rescue fund. We model nominal ROI with the rescue fund as:
\beq\label{eq: ROI^N alpha}
ROI_{i}^N(\alpha) = \frac{ROI^{int}\sum_{\substack{j\neq i}} A_{ij} + ROI^{ext}D_i+ ROI^E (1-\alpha)E_i+ROI^F\alpha E_i}{B_i}.
\eeq

Typical values of the parameters used in our simulations are $ROI^{int} = 0.04$, $ROI^{ext} = 0.07$, $ROI^E = 0.03$, $ROI^F = 0.02$ and $p^O=0.001$. We have thoroughly investigated other values of parameters, with special focus on the $p^O$, $ROI^E$ and $ROI^F$ finding the same qualitative picture in all cases. In some scenarios we find a declining linear regime in $ROI^{RA}$ for the cases when $ROI^F>ROI^E$ and $\alpha$ are relatively big. This is not surprising because for large $\alpha$ the cascade risk $p$ becomes very small and the leading order correction is then due to a change in $ROI^N(\alpha)$ as seen in~(\ref{eq: ROI^N alpha}). We present most of the results using aforementioned parameters as they are considered to be a reasonable estimate of real values based on the publicly available data on yield of specific asset classes~\cite{Schluter2012, iakova2001financial}. Note that in this analysis we assume total loss on the ROI after the default. Clearly, this assumption could be relaxed if needed.

\section*{Results}

Using the methodology reported in previous sections of this paper we run a number of simulations to investigate the relationships between systemic risk, risk-adjusted ROI, and the rescue fund. In figure~\ref{Fig: EtaP} we present the relationship between the level of balance sheet reserve requirements $\eta$ and the cascade risk $p^C$. Clearly, as the rate of reserve requirements increases, the cascade risk decreases. Also the cascade risk is much lower if some taxation rate $\alpha$ is present. The effect of reserve fund is especially strong for small and most realistic values of $\eta$. Different values in different panels are a consequence of different networks used for each panel. 

A more interesting question is how cascade risk changes with respect to taxation rate $\alpha$. The results are presented in figure~\ref{Fig: AlphaP}. It is obvious that the initial level of reserve requirements $\eta$ also strongly influences the outcomes. If there are no reserves, the rescue fund will not be funded regardless of the value of $\alpha$ and consequently the cascade risk does not depend on $\alpha$. As reserves increase, the parameter $\alpha$ plays more important role. The largest ameliorating effect of the rescue fund is found for the reserve level ranging between $0.001$ and $0.01$. The reserves to network exposure ratio in our model is correlated with the product $\beta\eta$, as seen in~(\ref{eq: Balance}) and~(\ref{eq: Reserve}). Therefore the ratio is most affected by $\alpha$ is roughly between $0.01$ and $0.1$. We believe that this effect depends on the details of the network and the chosen parameters. The sharp fall in the cascade risk for the small values of $\alpha$, points to the 
fact that the best strategy for the cascade risk mitigation is to combine both the reserve levels and the rescue fund.

It is also interesting to point out that according to our simulations, the cascade risk $p^{C}$ is lower for networks which were produced from the data for years 2008 and 2009, compared to networks produced from the period 2000-2004. It is possible that during the most intense period of 2007-2009 crisis banks traded with a higher level of systemic risk awareness and produced a more stable network of relationships themselves.

The DebtRank algorithm is also used to define a per node systemic risk measure called DebtRank. It signals the impact of the distress of an initial node across the whole network. The DebtRank of the node $i$, designated as $DR_i$, equals the fraction of the total economic value that is potentially under distress after the shock at node $i$. The average $DR$ is just the expected value of the DebtRank if initially shocked subject is chosen randomly (more details can be found in~\cite{battiston2012debtrank}). In the figure~\ref{Fig: EtaDR} we present the dependence of average DebtRank with respect to different levels of reserve requirements. It is not surprising that the average DebtRank is decreasing with an increase of reserves, but is interesting to note that for a very modest taxation rate of $\alpha=0.01$ the reduction of expected DR can be as high as $20 \%$. Overall the curves on this figure look very similar to the curves in figure~\ref{Fig: EtaP}.

In  figure~\ref{Fig: AlphaDR} we present the dependence between the taxation rate $\alpha$ and the DebtRank $DR$. The relationship is again similar to relationship between the taxation rate $\alpha$ and cascade risk $p$ presented in figure~\ref{Fig: AlphaP} and the conclusions are similar. Again, it is clear that the reserve level and taxation rate better work together to reduce the DebtRank of the system.

Aforementioned results indicate that a combination of the reserves and the taxation rate for rescue fund can greatly stabilize the system and reduce its systemic risk. As stated before, the proposed taxation strategy is devised so that it should not reduce the market liquidity, that it should be easy to implement and that it should not increase moral hazard in the system. In order to corroborate attractiveness of the proposed taxation scheme, we also compute the risk adjusted return on investments of the market players. 

In  figure~\ref{Fig: EtaROI} we show the dependence between $ROI^{RA}$ and the reserve level $\eta$. Interestingly, we find a function as monotonically increasing despite the effect that $\eta$ has on a nominal ROI. This feature is a consequence of chosen parameters because $E_i \ll B_i$ and the negative influence of the reserve on nominal ROI is small compared to its nonlinear influence on cascade risk. One could argue that this relation is in fact true as in reality financial institutions often hold more reserves then minimally required by imposed reserve requirements. The figure~\ref{Fig: AlphaROI} shows that the influence of taxation on $ROI^{RA}$ is not only strong but also beneficial for banks value. 

Finally in figure~\ref{Fig: IsoProbabilityCurves} we present iso-curves for the cascade risk. Iso-curves are obtained in the following way: First, a base reserve requirements $\eta_{0}$ is chosen. Then we compute cascade risk for each increase in reserve requirements $\frac{\eta-\eta_{0}}{\eta_{0}}$ and find $\alpha$ which would produce \emph{the same} value of cascade risk, \emph{if there was no increase in reserve requirements} ($\eta=\eta_{0}$), but only in the taxation for the rescue fund. It is clear that for all the chosen parameters, imposing a tax of $\alpha=1\%$ is much better than increasing $\eta$ for $1\%$. This is a final proof that in our model, in order to reduce systemic risk, introducing a rescue fund and taxing the reserves is more effective than increasing reserve or capital requirements. However, in the real world, the level of reserves impacts the probability of default $p^O$~\cite{agur2010capital}. This is a feature we did not model in our approach, but~\cite{agur2010capital} and~\cite{
thakor1996capital} points to strong negative effects of reserve requirements increase on the liquidity of the market. Since the rescue fund should not have such adverse effects we conclude that it is, at least, a viable idea to be tested by the regulators concerned with systemic risk.

\section*{Discussion}

Our model demonstrates that the cascade risk, as defined in this paper, is greatly reduced by even a small taxation for saving fund. The parameters used in simulation are artificial and there is a question of how well the DebtRank maps to the real world event of the default cascade. Nevertheless we made a case for simple proposed taxation strategy, as it is clearly more efficient in reduction of the cascade risk. Our research supports the following idea: If regulators aim at minimizing the systemic risk or more narrowly the cascade risk, the taxation for the fund is better on the performance level. Furthermore, although our implementation would require some modifications of the balance sheets, we argue that it is the cheapest of all proposed methods for the following reasons. Firstly, this allows to reduce the asymmetric impact of capital add-ons required to Global Systemically important financial institutions (G-SIFIs) introduced within the Basel III framework, which could corrupt the leveling the playing 
field principle. Secondly, 
it could help to design the reform for the super-national reform for safety nets, and the insurance deposit above all.

We have shown also that a very modest taxation can reduce the cascade risk so much that the risk adjusted ROI of individual subjects in a money market can greatly increase, thus making their coupons more profitable and thus possibly making them more responsive to presented idea. We believe that the Pigouvian nature of our tax and rescue fund would not increase moral hazard of individual subjects thus not affecting much the individual parameters, such as the individual probability of default.

With the introduction of the Pigouvian tax, the simulated contagion effect due to some bankrupt agents decreases to become "sustainable". The proposal sizes the rescue fund (i) big enough to cover the loss of most of the individual institutions in the market as a consequence of some failures; (ii) small enough to be easily and urgently replaced with the fresh liquidity once the fund has been used. In order to manage the stability – inefficiency trade off of financial firms, we paid a close attention to banks' profitability measuread with the return on investment (ROI) as well, because the regulatory measure has to both be efficient, and influence the market profitability as little as possible. A considerable portion of moral hazard is reduced by construction as the rescue fund assets do not cover initial troubled subject, but its counter parties. Accordingly with the profitability constraint, the impact of the reserve fund is strong with small and realistic values of the taxation rate. In order to minimize 
and manage the systemic risk, based on our analysis the tax could range from 2 to 4bps of banks' balance. To avoid any additional moral hazard behaviour, policy makers should better calibrate the tax taking into account all the money market trades to map more precisely the bank network topology and fixing a time period fixed tax.

In conclusion, the debate on rescue policies which ranges from raising of capital requirements to implementation of money market taxation on transactions is expected to enhance a fiscal proposal coherently with the economic theoretical background and the banking role played both in the monetary and credit markets.

\section*{Acknowledgements}
Both VZ and HA worked on the technical details of the paper. Survey of the literature was done by HA and GG and a mechanism for taxation was proposed by HA. All the authors contributed to writing of the paper. Authors acknowledge support from EU FP7 FET OPEN project FOC (Forecasting financial crises, grant no. 255987). VZ also acknowledges support from EU FP7 FET project MULTIPLEX (Foundational Research on MULTIlevel comPLEX networks and systems, grant No. 317532). GG also acknowledges support from EU Framework Program SSH.2010.1.2-1 (FESSUD) and SDA Bocconi School of Management for data acquisition. Authors also want to express their gratitude to Stefano Battiston, Giulia Iori and Hrvoje Stefancic on fruitful comments of the paper. 

\medskip

\bibliography{biblio}

{\bf
\section*{Figure Legends}
\newpage
\begin{figure}[ht!]
\begin{center}
\includegraphics[width=0.8\textwidth]{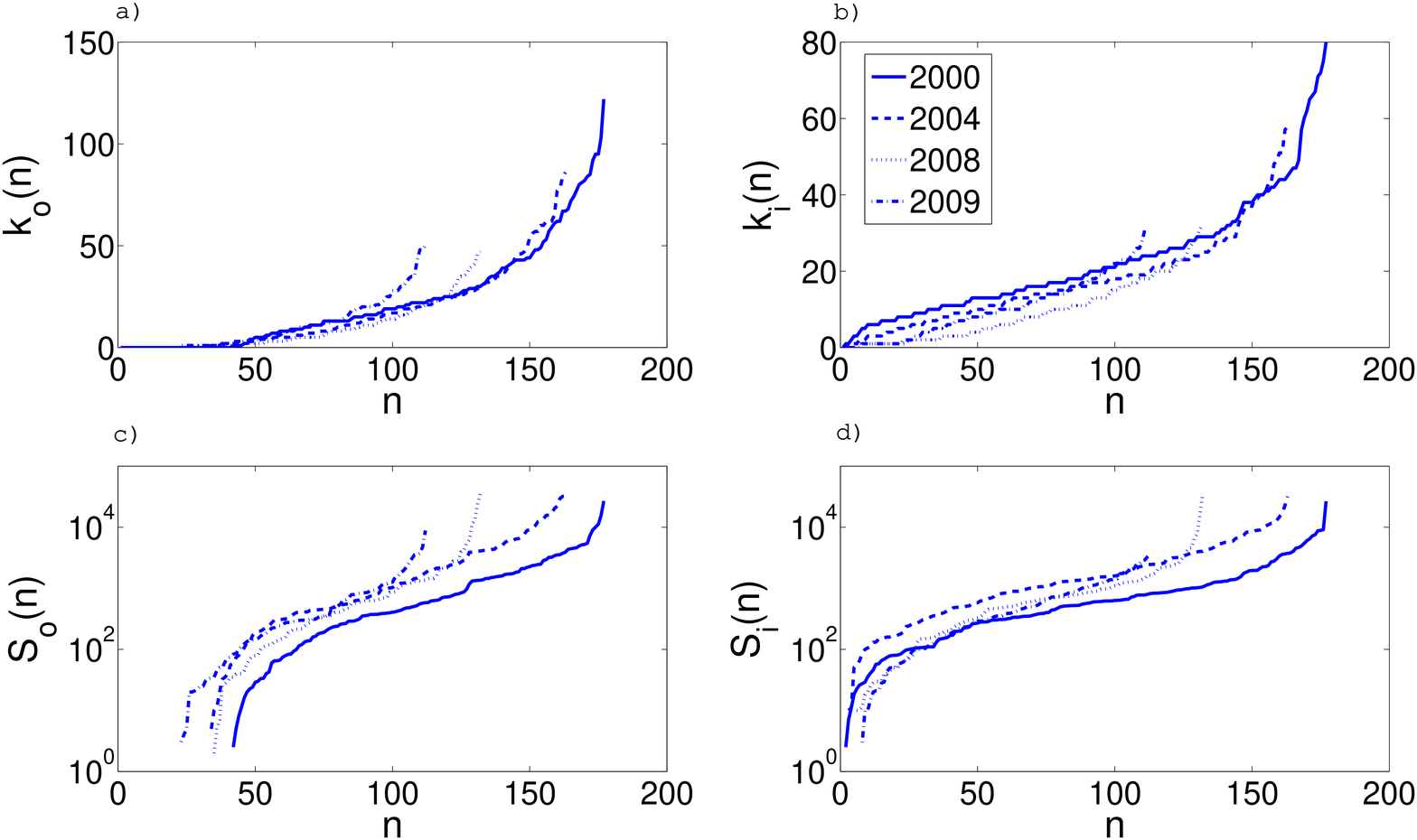}
\end{center}
\caption{{\bf In this picture x-axis is the order of the node with respect to its value on the y-axis.} The panel a) has out-degree on the y-axis, the panel b) has in-degree on the y-axis, the panel c) has out-strength on the y-axis and the panel d) has in-strength on the y-axis. The heterogeneity of all the properties is clear.}
\label{Fig: NetworkProperties}
\end{figure}

\begin{figure}[ht!]
\begin{center}
\includegraphics[width=0.8\textwidth]{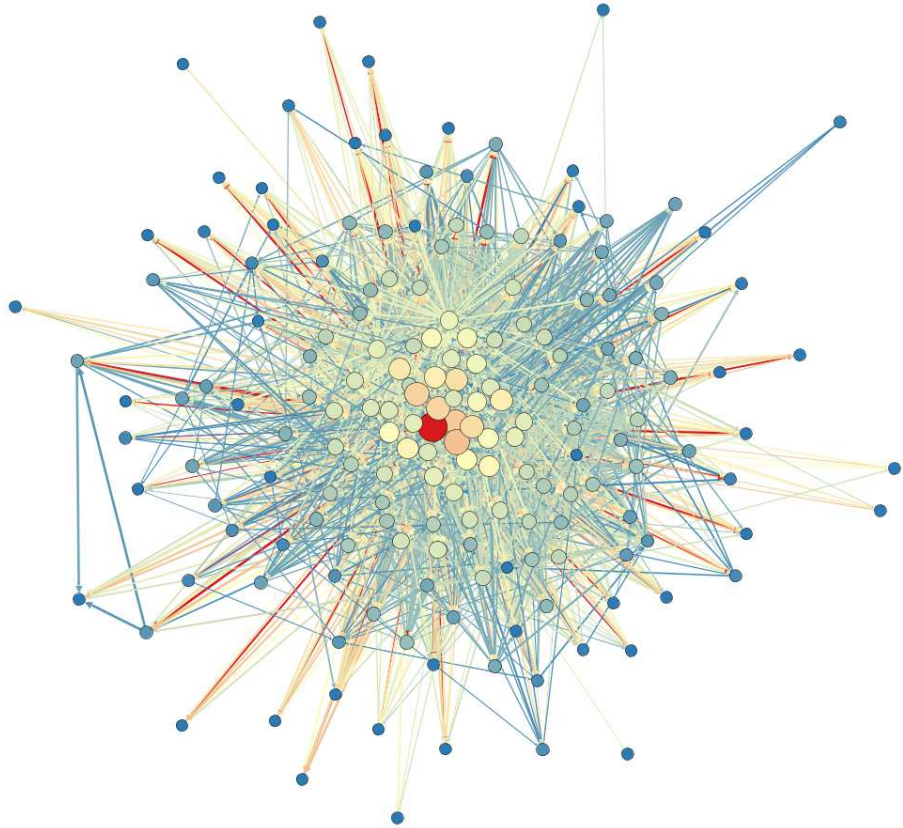}
\end{center}
\caption{{\bf This figure shows a network of e-MID precipitants for April 2000.} Thicknesses of edges $A_{ij}'$ are recalculated from the real weights $A_{ij}'=log(A_{ij}/(0.25\cdot A_{min}))$. The core-periphery structure and weight heterogeneity are easily observed. Size and color of the nodes correspond to out strength of the node computed with rescaled weights. Palette is color coded so that the blue links (nodes) have the lowest weight (out strength) while red carry the largest weight (out strength).}
\label{Fig: Network}
\end{figure}

\begin{figure}[ht!]
\begin{center}
\includegraphics[width=0.8\textwidth]{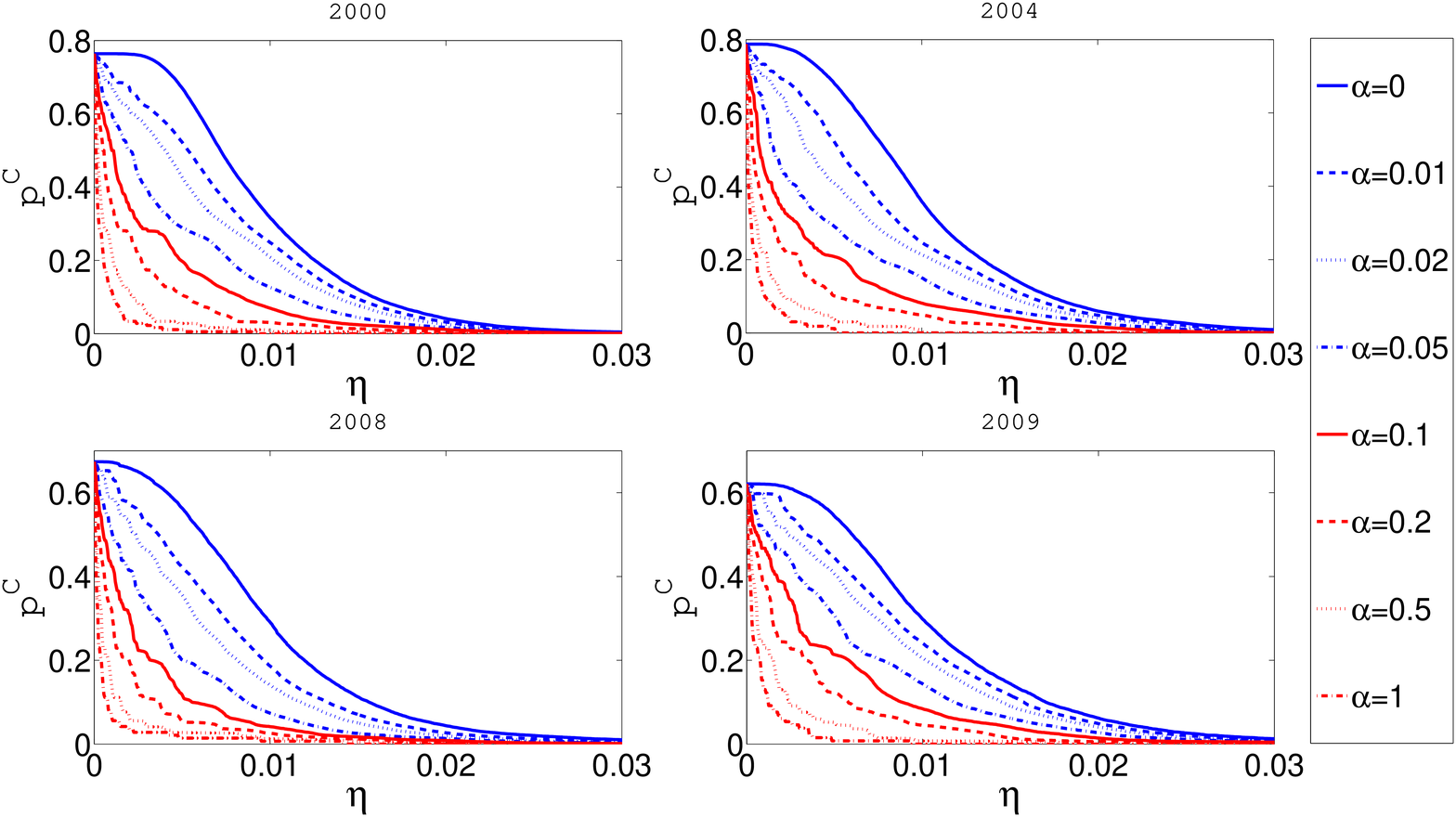}
\end{center}
\caption{{\bf In this figure x-axis represents $\eta$ and y-axis represents cascade risk $p^C$.} Different panels represent the results for networks obtained from the aggregate of some one month data for four different years. Fixed taxation rate $\alpha$ is constant for every curve in the plot.}
\label{Fig: EtaP}
\end{figure}

\begin{figure}[ht!]
\begin{center}
\includegraphics[width=0.8\textwidth]{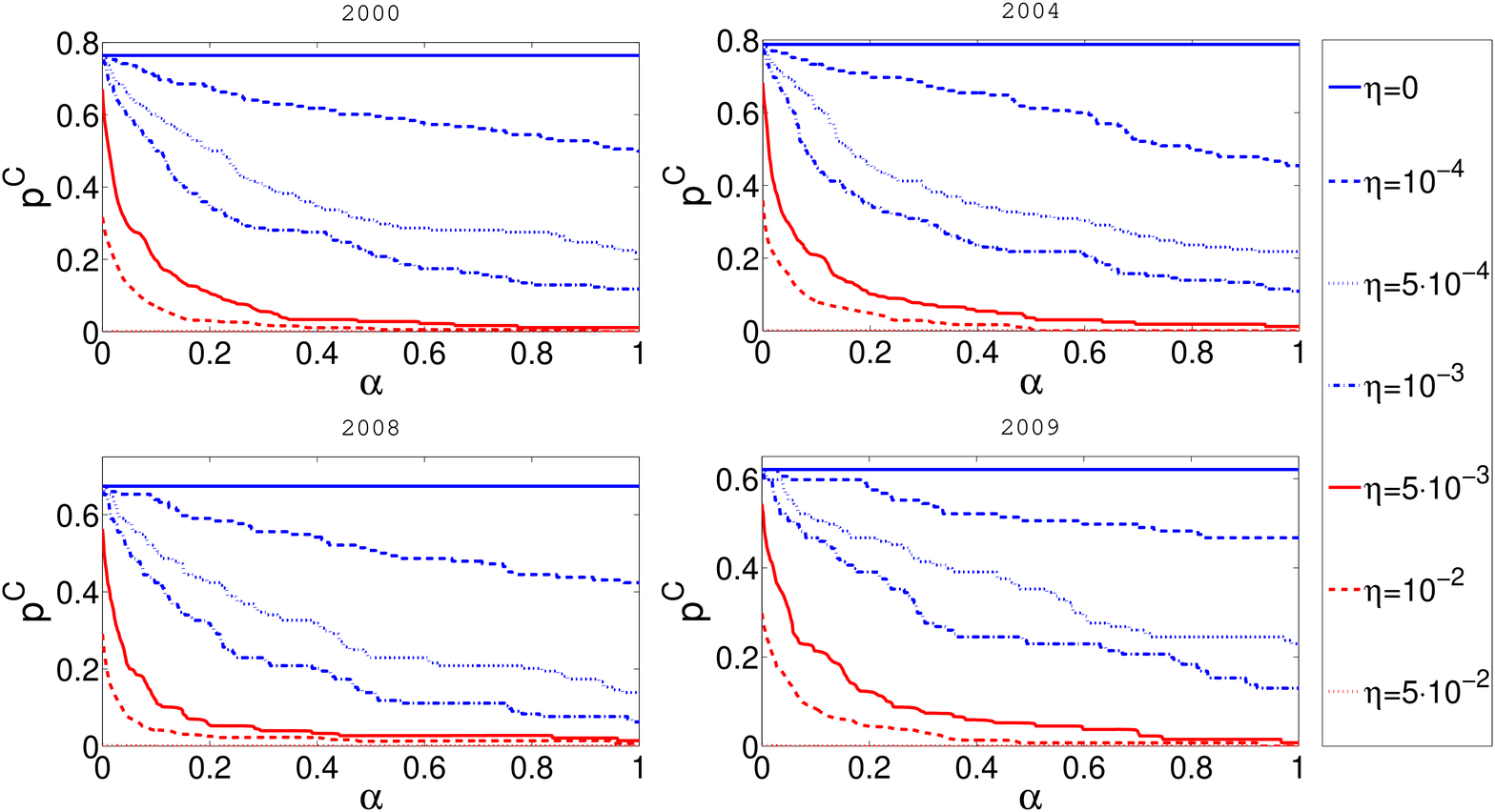}
\end{center}
\caption{{\bf In this figure on the x-axis represents $\alpha$ and y-axis represents cascade risk $p^C$.} Different panels represent the results for networks obtained from the aggregate of some one month data for four different years. Balance sheet reserve requirement rate $\eta$ is constant for every curve in the plot.}
\label{Fig: AlphaP}
\end{figure}

\begin{figure}[ht!]
\begin{center}
\includegraphics[width=0.8\textwidth]{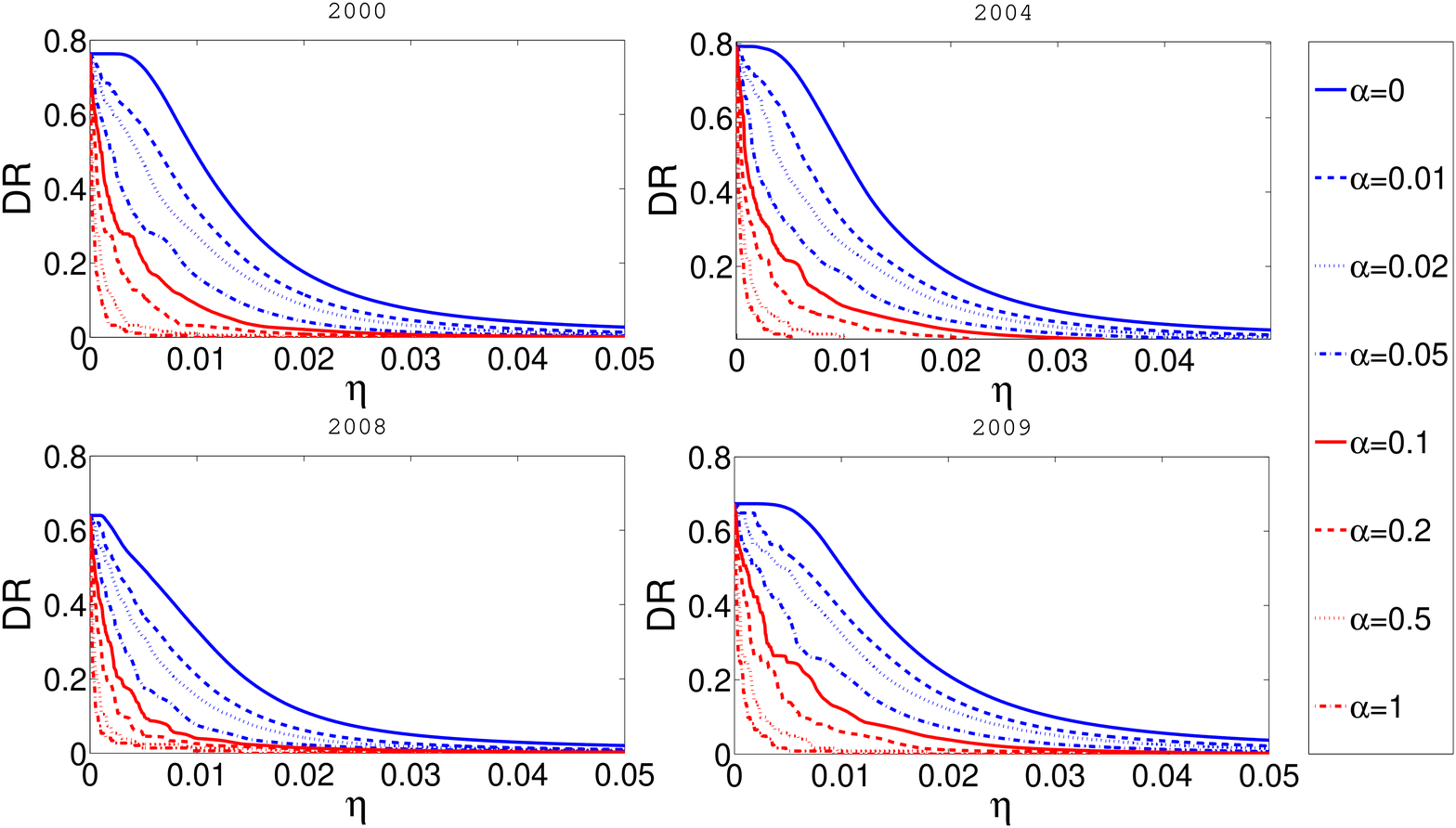}
\end{center}
\caption{{\bf In this figure x-axis represents $\eta$ and y-axis represents systemic DebtRank $DR$.}  Different panels represent the results for networks obtained from the aggregate of some one month data for four different years. Fixed taxation rate $\alpha$ is constant for every curve in the plot.}
\label{Fig: EtaDR}
\end{figure}

\begin{figure}[ht]
\begin{center}
\includegraphics[width=0.8\textwidth]{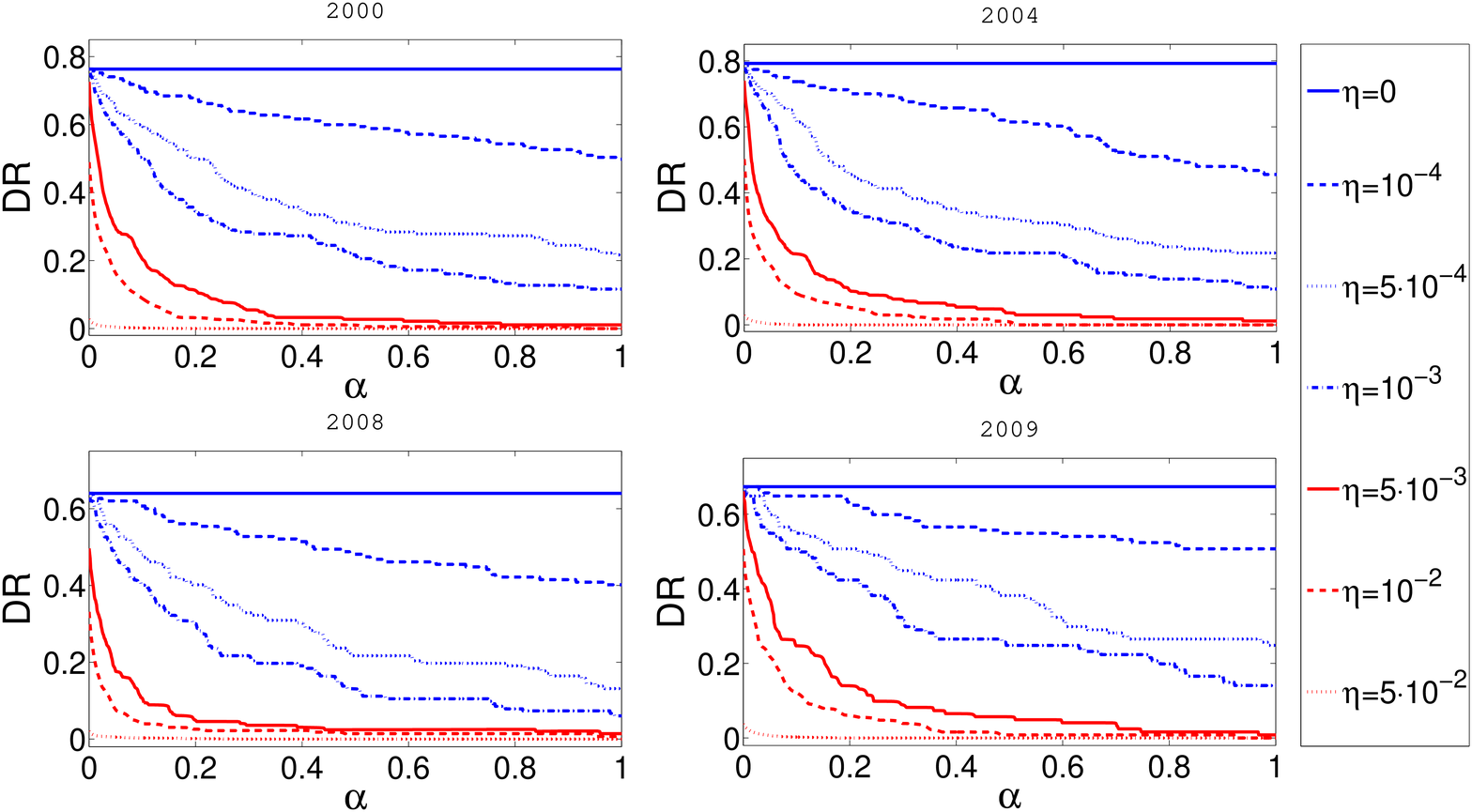}
\end{center}
\caption{{\bf In this figure x-axis represents $\alpha$ and y-axis represents systemic DebtRank $DR$.} Different panels represent the results for networks obtained from the aggregate of some one month data for four different years. Balance sheet reserve requirement rate $\eta$ is constant for every curve in the plot.}
\label{Fig: AlphaDR}
\end{figure}

\begin{figure}[ht!]
\begin{center}
\includegraphics[width=0.8\textwidth]{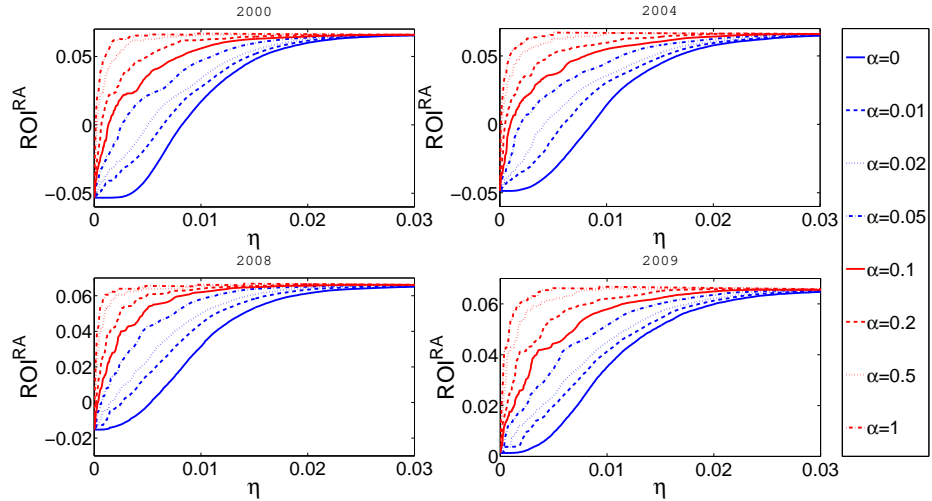}
\end{center}
\caption{{\bf In this figure x-axis represents $\eta$ and y-axis represents risk adjusted return on investment $ROI^{RA}$.} Different panels represent the results for networks obtained from the aggregate of some one month data for four different years. Fixed taxation rate $\alpha$ is constant for every curve in the plot.}
\label{Fig: EtaROI}
\end{figure}

\begin{figure}[ht!]
\begin{center}
\includegraphics[width=0.8\textwidth]{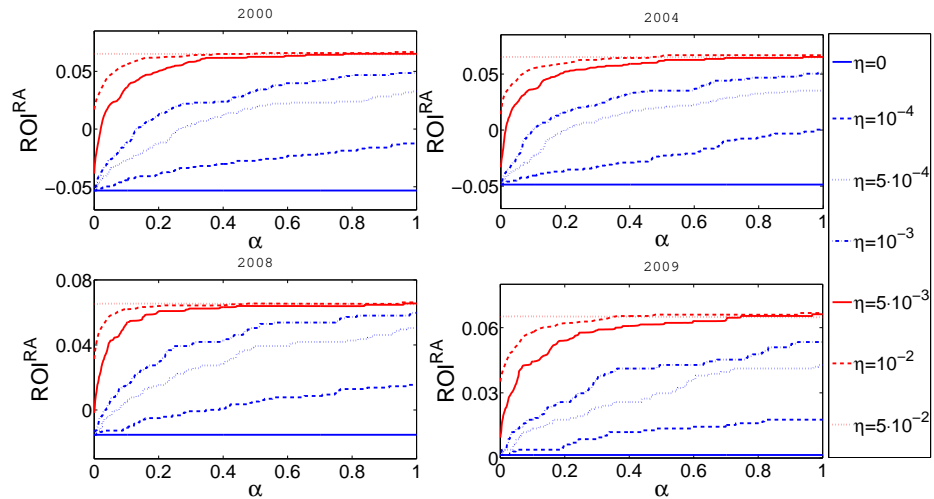}
\end{center}
\caption{{\bf In this figure x-axis represents $\alpha$ and y-axis represents the risk adjusted return on investment $ROI^{RA}$.} Different panels represent the results for networks obtained from the aggregate of some one month data for four different years. Balance sheet reserve requirement rate $\eta$ is constant for every curve in the plot.}
\label{Fig: AlphaROI}
\end{figure}

\begin{figure}[ht!]
\begin{center}
\includegraphics[width=0.8\textwidth]{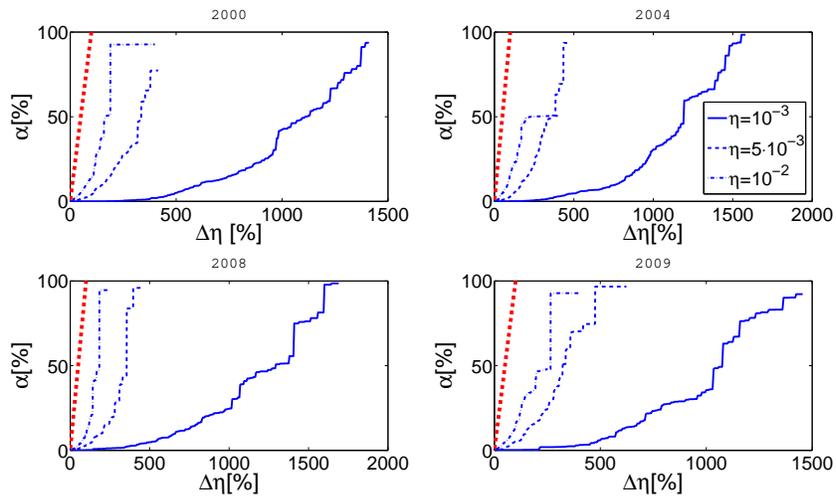}
\end{center}
\caption{{\bf In this figure x-axis represents percentage of change of $\eta$ and y-axis represents $\alpha$.} Curves represent the points at which the systemic probability of default takes the same value in the case \emph{(i)} when no fund is present, i.e. $\alpha=0$ and the system is stabilized through increase of reserve requirements by $\Delta\eta$ and \emph{(ii)} when there is no change in reserve requirements, but the stabilization of the system is implemented through the rescue fund. Three starting levels of reserve requirements are presented through three different curves. The red fat dashed line represents hypothetical case in which the influence of reserve requirements increase by percentage $x$ would be exactly the same as if the proposed taxation was implemented with the same rate $x$. Different panels represent the results for networks obtained from the aggregate of some one month data for four different years.}
\label{Fig: IsoProbabilityCurves}
\end{figure}

\end{document}